\documentclass[epj-spec]{svjour}
\usepackage{graphicx}
\begin{document}
\title{Phase-Space Coalescence for heavy and light quarks at RHIC}
\author{V.Greco\inst{1}\fnmsep\thanks{\email{greco@lns.infn.it}}}
\institute{Dipartimento di Fisica e Astronomia, Via S. Sofia 64, I-95125 Catania, Italy  \and INFN-LNS, Via S. Sofia 62, I-95125 Catania, Italy}
\abstract{We review the application and successes of a phase-space coalescence plus fragmentation model that
has been applied for hadronization at RHIC.
The physical concept is discussed together with the practical implementation.
The robustness of main predictions is reviewed together with several open issues like
relevance of three dimensional calculation, finite width of the wave functions, effects
of quark masses, energy-entropy conservation, space-momentum correlation. 
Eventually the relevance of coalescence also for the study of the
microscopic interaction of heavy quarks is highlighted.
} 
\maketitle
\section{Introduction}
\label{intro}
The goal of ultra-relativistic heavy-ion collisions is to create a new state
of nuclear matter with quarks and gluons as degrees of freedom.
Such a state of matter is predicted to be one of the Quantum Chromodynamics
(QCD) phases, usually referred to as quark-gluon plasma (QGP), and it is expected to
 occur at energy densities above $\epsilon_c \simeq 0.7$ GeV/fm$^3$ 
and temperatures above $T_c \sim$ 180 MeV \cite{lattice}.
These conditions are believed to be reached at the Relativistic Heavy Ion
Collider (RHIC) as confirmed by the success of ideal hydrodynamics \cite{Heinz:2004pj,kolb-heinz} (at low $p_T <$1.5 GeV)
and the observed strong nuclear suppression $R_{AA}$ in agreement with
jet-quenching predictions \cite{gyulassy1}. In fact both entail an energy density 
$\epsilon \geq10 $ GeV/fm$^{3} \, \gg \epsilon_c$.  

Such a hot and dense medium is therefore expected to be in a deconfined state,
i.e. a state where quarks and gluons act directly as degrees of freedom.
The success of hydrodynamics based on a QGP equation state in reproducing
the observed large elliptic flow \cite{Heinz:2004pj,kolb-heinz}, 
together with the failure of hadronic transport model \cite{bleicher},
constitutes a first indirect proof of the partonic phase.
However in the search for signatures of the QGP and in the study of
its properties an important role should be played by the understanding of hadronization dynamics,
as persistently stressed by J. Zimanyi \cite{Zimanyi:2005nn}.
However because of its intrinsic non-perturbative
nature there has been a general skeptical view on the real possibility
to have hints of quark degrees of freedom. 
Surprisingly
the wide availibility of exclusive data from RHIC experiments and the
power of the phenomenological approach have made
possible to spot signatures of hadronization via coalescence of massive quarks
\cite{hwa,greco,greco2,greco-c,fries,moln}.

Coalescence is a quite general concept that has been applied to nuclear physics
for light cluster production for more than two decades \cite{coal-hadr}. 
First application to ultra-relativistic heavy-ion
collisions is due again to J. Zimanyi and his group
in Budapest who implemented an ALgebraic COalescence Rehadronization (ALCOR) model
that studies the chemical composition of the hadronic matter produced in heavy ion 
collisions at SPS and RHIC \cite{alcor}. However it was only
in 2002-03 that some puzzling behavior of the experimental data
in Au+Au collisions at RHIC showed features specific of
valence quark coalescence: enhancement of baryon/meson ratio at intermediate $p_T$,
absence of $R_{AA}$ suppression for baryons,
and the scaling of elliptic flow 
according to the costituent quark number \cite{greco,greco2,fries,moln,voloshin}.

In this talk, we will review the general idea of coalescence and its phase-space
implementation for the hadronization. Then we will show the 
comparison with the experimental data for the particle spectra, their 
ratios and elliptic flows. Finally we will discuss the robustness of the basic
feautures of a naive coalescence approach together with some open issues.
Quark coalescence was initially applied to light quarks, but I will also draw the attention to
its role in the investigation of the in-medium interaction of heavy quarks and quarkonia in the 
QGP \cite{greco-c,vanHees:2004gq,vanHees:2005wb}.

\section{Coalescence and Fragmentation}
\label{fra-coal}
A lot of efforts has been done to work on the hadronization process, even
if a fully satisfying theoretical description is still missing. 
Because of its highly non perturbative nature, 
hadronization is treated through different phenomenological approaches
that have shown their validity in different momentum scale regions:
string fragmentation, indipendent fragmentation, dual parton model, 
coalescence. 
We will discuss mainly a coalescence plus fragmentation model
that has the advantage to describe in an economical way different
features of the hadronization and the collective behavior of the hadronic matter created at RHIC.

We briefly remind the main features of indipendent fragmentation (IF) used
as standard hadronization mechanism for enough large $p_T$.
IF relies on the QCD factorization theorem that
allows to write the hadron production as a convolution 
$f_H(p_T)=\sum_p f_p(p_T/z) \otimes D_{p \rightarrow H}(z,\mu)$, where
$f_p$ is the parton distribution in the system, $D_{p\rightarrow h}$ is the
fragmentation function, with
$z$ fraction of the momentum of the parton carried by the hadron 
and $\mu \sim p_T^2$ the pertubative scale.
In  $e^+ e^-$ and $hh^\prime$ collisions, $f_p$ can be determined by the pQCD cross section folded
with the parton distribution functions (PDF's), while 
in heavy-ion collisions they are calculated as sovrapposition of $pp$ collision
plus nuclear shadowing and, moreover, radiative energy loss 
due to in-medium  gluon-radiation \cite{gyulassy1}.
In such an approach non-pertubative features are encoded in the fragmentation function $D_{p \rightarrow H}(z,\mu) $, that gives the probability to produce from a parton $p$ a hadron 
with momentum $p_h=z p_T $. Such a function is considered to be universal 
and is
extracted from the available data in $e^+e^-$ collisions and then applied to study
also hadron-hadron collisions. This approach has been successful in reproducing
the $\pi^0$ transverse momentum distribution down to $p_T \sim$ 2 GeV \cite{pp-frag} for
$pp$ collisions at RHIC energy.

Despite the success of IF for high energy collisions and hadrons at large $p_T$,
at lower energy or large rapidity there have been evidences of hadronization mechanism
that proceeds through a coalescence mechanism. For example
the $D^-/D^+ $ ratio in hadron-nucleus reaction at Fermilab \cite{Rapp:2003wn,braa} or
the particle production in the fragmentation region (large rapidity)
that cannot be explained by IF while coalescence seems a more suitable approach
\cite{hwa-tfr}. In such physical conditions it seems necessary to postulate that hadrons
come from a convolution of two parton distribution folded by a wave function
$f_H(p)= f_p(p_{1}) \otimes f_p(p_{2}) \otimes \Phi_M (p_1-p_2)$,
where $p_1$ and $p_2$ are the momenta of coalescing partons.
One naively expects that when the phase space is enough dense and the process
involves a low virtuality, the production of quarks from vacuum is 
not likely and the recombination of quarks
becomes the dominant mechanism of hadronization. 
In this perspective HIC's provide the ideal environment much denser than $hp$
collisions in a wider kinematical region (respect to $hh^{\prime}$ collisions), 
for which is natural to think that recombination processes may
play a dominant role, at least at not much high $p_T$.

On the other hand it is quite natural to think that nature has a smooth
transition between the two processes as a function of phase space and virtuality,
hence IF or, at lower $p_T$, string fragmentation could always be present together
with coalescence.

\section{Phase -Space Formulation and Implementation}
\label{formulation}
We describe our implementation of coalescence here, while the fragmentation part is described
in the next subsection. Our approach is based on the Wigner formalism \cite{coal-hadr} that allows
a more direct connection with the dynamical phase-space description of heavy-ion collisions
(HIC's).
In this formalism the transverse momentum spectrum of 
hadrons that consist of $n$ (anti-) quarks  is given by the
overlap between the hadron wave function and the $n$ quark phase-space distribution function 
$f_{q}(x_{i},p_{i})$:

\begin{equation}
\frac{dN_{H}}{d^2P_T}=g_{H}\int \prod_{i=1}^{n}\frac{d^{3}\mathbf{p}_{i}}
{(2\pi)^{3}E_{i}}{p_{i}\cdot 
d\sigma _{i}}f_{q}(x_{i},p_{i}) f_{H}(x_{1}..x_{n};p_{1}..p_{n})\,\delta^{(2)}
\left(P_T - \sum_{i=1}^n p_{T,i}\right)
\label{coal1}
\end{equation} 

where $d\sigma$ denotes an element of a space-like hypersurface,
$f_H(x_{1}..x_{n};p_{1}..p_{n})$ is the Wigner distribution function of the 
hadron, $g_H$ is the probability of forming from $n$ colored quarks 
a color neutral object with the spin of the hadron considered. In Eq.(\ref{coal1})
it is already assumed that the $n$ quark phase space distribution is approximated by 
the product of the single quark distribution function
\begin{equation}
 f_q(x_{1}..x_{n};p_{1}..p_{n})=\prod_{i=1}^{n} f(x_i,p_i)
\end{equation}
and therefore no quark-quark correlations are included. The formalism can be formally
extended to include such correlations \cite{Fries:2004hd}, but then it is necessary a dynamical
study of the in medium correlations that asks for transport approaches.

As light hadrons wave function we have used a sphere in both space and momentum,  
with radii $\Delta_r$ and 
$\Delta_p$, respectively, which in the Wigner formalism
are related by $\Delta_r\cdot\Delta_p=1$.
A good description of pion, kaon, proton, antiproton spectra can be obtained with 
a radius parameter $\Delta_p= 0.24$ GeV for mesons and $0.36$ GeV 
for baryons, which in terms of mean square radius corresponds to take a
slightly larger radius for baryons respect to mesons. 
The multidimesional integral Eq.(\ref{coal1}) is evaluated
in the full 6D phase space
by the Monte Carlo method via test particle method \cite{greco2}.

\subsection{Bulk properties}
For partons in the quark-gluon plasma we take a thermal distribution 
for transverse momenta up to $p_0 = 2$ GeV in agreement
with the hydrodynamical behavior observed at low $p_T$.
For their longitudinal
momentum distribution we assume boost-invariance, i.e.
a uniform rapidity distribution in the rapidity range $y\in(-0.5,+0.5)$.
To take into account collective flow of quark-gluon plasma, these
partons are boosted by a flow velocity 
${\bf v}_{\rm T}=\beta_0({\bf r}_{\rm T}/R)$, 
depending on their transverse radial position $r_{\rm T}$,
$R$ being the transverse size of the quark-gluon plasma at hadronization, 
and $\beta_0$ is the maximum collective flow velocity of the quark-gluon plasma.
Therefore for light quarks and antiquarks transverse
momentum spectra are given by 
\begin{eqnarray}\label{quark}
\frac{dN_{\rm q,\bar q}}{d^2{\bf r}_{\rm T}d^2{\bf p}_{\rm T}}
=\frac{g_{q,\bar q}\tau m_{\rm T}}{(2\pi)^3}
&\times&\exp\left(-\frac{\gamma_{\rm T}(m_{\rm T}-{\bf p}_{\rm T}\cdot
{\bf v}_{\rm T}\mp\mu_q)}{T}\right),
\end{eqnarray}
where $g_q=g_{\bar q}=6$ is the spin-color degeneracy of light quarks
and antiquarks, and the minus/plus signs
of chemical potentials are for quarks and antiquarks, 
respectively. The slope parameter $T$ is taken to be $T=170$ MeV, 
consistent with the phase transition temperature 
from lattice QCD calculations \cite{lattice}.

The masses of thermal quarks are taken to be those of constituent quarks, 
 $m_{u,d}=300$ MeV, $m_s=475$ MeV, $m_c=1.4$ GeV and $m_b=4.8$ GeV
(effect of masses on the typical observed scaling is discussed in Section \ref{light}). 
We notice that in such an approach non perturbative effect are also encoded in the 
quark masses, in fact a large part of the interaction can be accounted for
by quark thermal masses \cite{levai-eos,Castorina:2005wi,Castorina:2007qv};
however the relation of this masses to the masses associated to scalar condensates
remains still an open challenging theorethical question.
For the quark chemical potential $\mu_q$, we 
use a value of $\mu_q=10$ MeV to give a light antiquark to quark ratio of 
0.89, which would then lead to an antiproton to proton ratio of about 
$(0.89)^3=0.7$, consistent with the observed ratio at midrapidity in heavy 
ion collisions at RHIC.

Above $p_0$, partons are from the quenched pQCD minijets \cite{levai1}, and 
their masses are those of current quarks (which schematically resembles the 
$p-$dependence of the constituent quark masses).  
Both soft thermal and hard minijet partons are assumed 
to be distributed uniformly in a fireball \footnote{Such an assumption is in principle too drastic for minijets, however it does not affect our results 
because jet-jet coalescence is never taken into account.} 
of volume of 950 
fm$^3$ (which implies a transverse radius $R= 8.1$ fm and longitudinal length of $\tau= 4.4$ fm
for a Bjorken plus transverse expansion). Such a volume is fixed to reproduce the measured 
transverse energy of 750 GeV 
\cite{bazi}, 
together with radial flow parameter $\beta_0= 0.5$  which is consistent with both experimental data 
and hydrodynamical calculations \cite{kolb-heinz}.
In other words the reconstruction of the fireball
tells us that the system used for coalescence hadronizes at an energy density of 
0.8 GeV/${\rm fm^3}$, which is essentially the energy density at which the
phase transition is expected to occur from lattice QCD calculations 
\cite{lattice}.
In addition the entropy of such a system is $dS/dy\cong 4800$ in agreement with the value
inferred from experimental data by S. Pratt and S. Pal \cite{Pal:2003rz}.
Therefore even if employed mainly at intermediate $p_T$ coalescence model relies on a bulk
distribution that is consistent with what can be inferred
from hydrodynamics and experimental data.

\subsection{Minijet Distribution}
The transverse momentum distribution for $p_T > 2$ GeV is taken to be that of minijet partons 
in the midrapidity. In particular we use the one that can be obtained from an improved 
perturbative QCD calculation \cite{yi02}. 
It is given by $dN_{\rm jet}/d^2p_T=1/\sigma^{0-10}_{\rm tot} 
d\sigma_{\rm jet}/d^2p_T$ in terms of $\sigma^{0-10}_{\rm tot}$ corresponding
to the total cross section at central 10\% of the collisions and the
jet production cross section from nucleus-nucleus collisions,
\begin{eqnarray}
\frac{d\sigma_{\rm jet}}{d^2{\bf p}_{\rm T}}\ & =& \ 
\int d^2{\bf b}\ d^2{\bf r}\ t_{\rm Au}({\bf r}) t_{\rm Au}({\bf b}-{\bf r}) 
\sum_{\!\!ab}\!\!\int\!\! dx_a dx_b d^2{\bf k_a}_{\rm T} 
d^2{\bf k_b}_{\rm T} g({\bf k_a}_{\rm T}) g({\bf k_b}_{\rm T})\nonumber \\
&\times&f_{\!a/{\rm Au}}(x_a,Q^2)f_{\!b/{\rm Au}}(x_b,Q^2)
\frac{\hat s}{\pi}\delta(\hat s+\hat t+\hat u)
\frac{d\sigma^{ab}}{d\hat t}.
\end{eqnarray}
In the above, $t_{\rm A}({\bf r})$ is the thickness function of Au at 
transverse radius ${\bf r}$. The parton distribution 
function in a nucleon in the nucleus Au is denoted by 
$f_{a/{\rm Au}}(x,Q^2)$ plus a transverse smearing $g({\bf k}_{\rm T})$. 
The cross section $d\sigma^{ab}/d\hat t$ is the parton scattering
cross section. Kinematic details and a systematic analysis of $pp$
collisions can be found in Ref.~\cite{yi02}. Using the GRV94 LO result 
for the PDF \cite{structure} and the KKP fragmentation function from 
Ref. \cite{KKP}, measured data in the reaction $pp\to\pi^0X$ at 
$\sqrt{s}=200$ GeV can be reproduced with $Q=0.75 p_{\rm T}$ and 
$\langle k_{\rm T}^2 \rangle = 2$ GeV$^{2}$. 

In heavy ion collisions at RHIC, minijet partons are expected to lose 
energy by radiating soft partons as they traverse through the quark-gluon 
plasma \cite{gyulassy1}. This effect is taken into account by lowering their transverse
momenta by the energy loss $\Delta E$, which depends on both the parton
energy $E$ and an effective opacity parameter $L/\lambda$ according to 
the GLV model \cite{levai1}. An 
effective opacity $L/\lambda=3.5$ is used, as extracted from a fit to the spectrum 
of high transverse momentum pions measured at RHIC \cite{levai1} \footnote{In this
calculation is not included the QCD analog of the Ter-Mikayelian effect. When
included a $L/\lambda=5$ has to be used to have the same amount of quenching \cite{djord}.}.

We note that our approach is indeed quite schematic in separating the thermal
spectrum from the one of minijets. In reality there is only one spectrum
with possible correlations also at $p_T < p_0$. Our simplificaton does not 
affect the one-body observables described in the following, but can be 
essential for studying two or three particle angular correlations \cite{phenix-corr,star-corr}.

\section{Coalescence for light quarks}
\label{light}
\begin{figure}
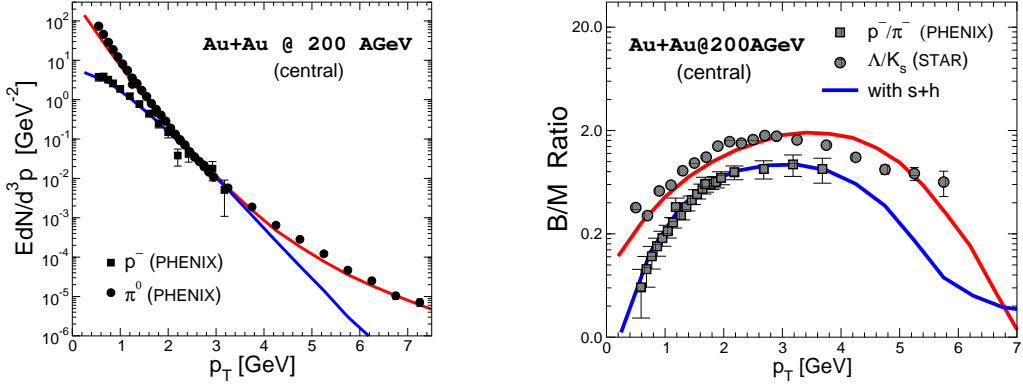

\begin{center}
\includegraphics[height=2.in]{pion-aprot-spec.eps}
\hspace{0.5in}
\includegraphics[height=2.in]{ratio-pip-kl.eps}
\end{center}
\caption{Left: Pion and antiproton transverse momentum spectra 
from Au+Au collisions at 
$\sqrt s=200$ AGeV, calculation from coalescence plus fragmentation of minijets
is shown by solid lines; experimental data \cite{phenix-spec} for  $\pi^0$ are shown by 
filled circles, and for $\bar{p}$ by squares.
Right: $\bar{p}/\pi$ and $K^0_s/\Lambda$ ratios, model by solid lines and
experimental data \cite{ppi,Lamont:2004qy} by filled squares.}
\label{spectra}
\end{figure}

In the coalescence model, hadrons are formed from quarks that are close 
in phase space if a hadron wave function with a small width is considered. 
As a result, baryons with momentum $p_T$ are produced 
from quarks with momenta $\sim p_T/3$, while mesons with same 
momentum are from quarks with momenta $\sim p_T/2$. Since the
transverse momentum spectra of quarks decrease with $p_T$,
production of high momentum baryons from quark coalescence is 
favored respect to the fragmentation where baryons are penalized with respect to  
mesons as more quarks are needed from the vacuum. In the independent fragmentation
the estimated value of $p/\pi$ ratio is of $\sim 0.2\div0.3$ in approximate agreement
with the ratio measured in $pp$ collisions.
For Au+Au collisions results for $\bar p$ and $\pi^-$ spectra are shown (solid lines) together with the data
from PHENIX \cite{phenix-spec} in Fig.\ref{spectra}(left). 
The resulting $\bar p/\pi$ and the $K_s^0/\Lambda$
ratios based on Eq.(\ref{coal1}) are shown in Fig.\ref{spectra} (right) 
together with data points \cite{esumi,Lamont:2004qy}. 
Calculations shown for $\pi,\, K$ include the constribution from indipendent fragmentation
that is essential to reproduce the meson spectra already at $p_T \sim 3$ GeV (especially for pions).
Although different models have been used, they all lead to enhanced $\bar p/\pi$ ratio and $K_s^0/\Lambda$ ratio
of similar magnitude \cite{hwa,fries}.
As shown in Fig.\ref{spectra} (left), our approach \cite{greco,greco2},
which includes resonance decays and avoids the collinear approximation by using the 
Monte Carlo method to evaluate the multi-dimensional coalescence 
integral, gives a good description of spectra, $\bar p/\pi$ and $K_s^0/\Lambda$
ratios also at $p_T < 2$ GeV, even if in this momentum region the assumptions used 
are less under control (for a more extended discussion see next Section).
At lower energy, Au+Au at 62 AGeV, a direct extrapolation with the same
model predicts \cite{Greco:2005jk} an enhancement of
the $p/\pi$ ratio and a decrease of $\bar{p}/\pi$ respect to 200 AGeV. Such a trend 
is again in reasonable agreement
with experimental data \cite{Mohanty:2007ru,Abelev:2007ra}.

In HICs the anysotropy of  particle momentum distributions
with respect to the azimuthal angle offers the possibility
to get information on the dynamics of the collisions and the properties of the
created hot and dense matter \cite{Lacey:2006pn}. At RHIC energies it has been
shown to be an important probe for the equation of state of the 
quark-gluon plasma\cite{kolb-heinz} and of the parton cross section \cite{zhang,moln02}, 
and for hadrons
at high transverse momentum it is a probe of the initial energy density
that causes the jet quenching \cite{gyulassy1}.
Moreover it has been shown that it is built-in very early in the 
dynamical evolution being thus a probe of the interaction during the partonic
stage \cite{zhang}, a picture recently confirmed by the measurement of $\phi$ meson $v_2$ \cite{Afanasiev:2007tv}.

\begin{figure}
\begin{center}
\includegraphics[height=1.9in]{v2-ppi-fin.eps}
\hspace{0.5in}
\includegraphics[height=2.0in]{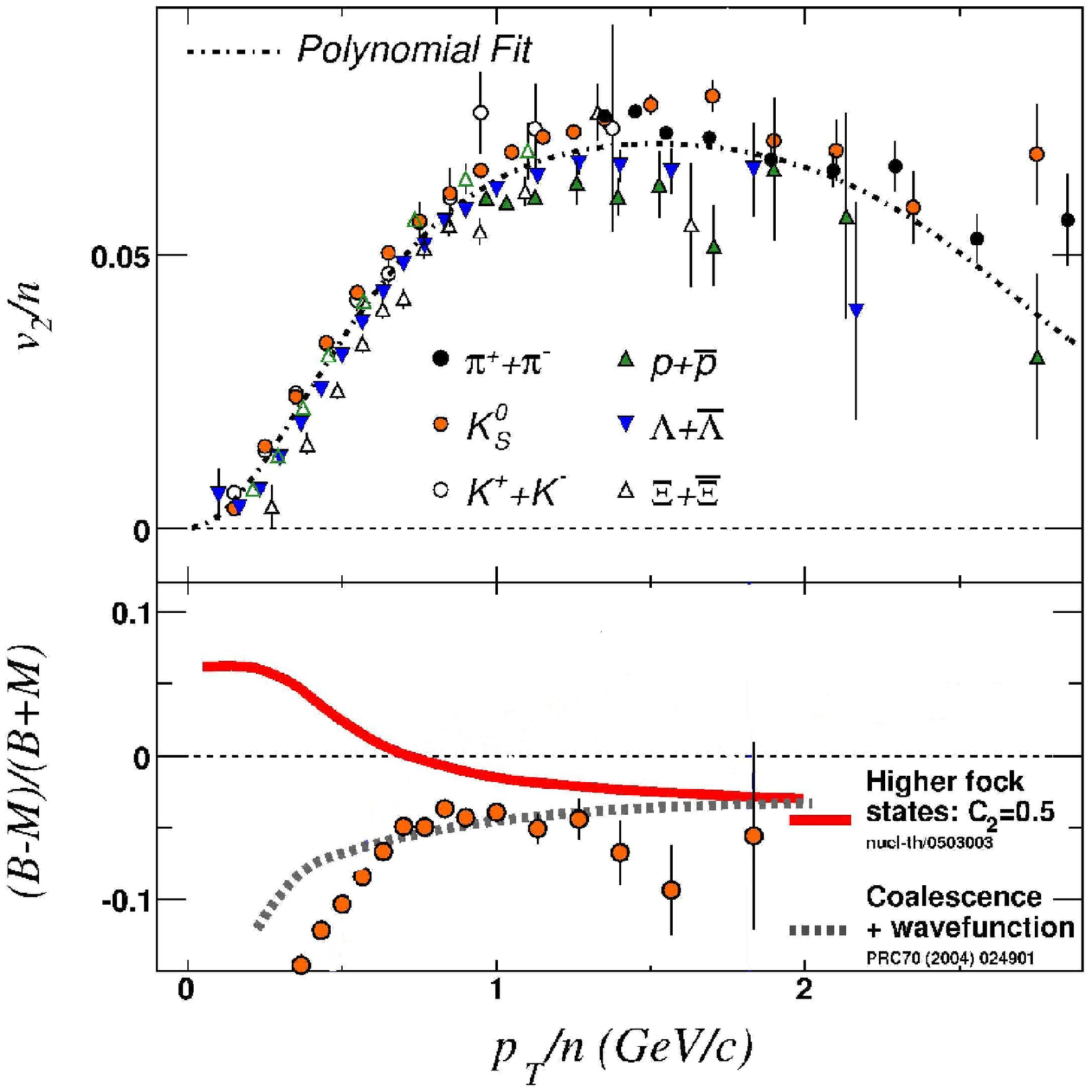}
\end{center}
\caption{Left: quark number scaled elliptic flows of $\pi$ and $p$ in Au+Au at 200 AGeV. 
Lines are from the coalescence model, while symbols are data from PHENIX \cite{adlv2}. Right: 
(top) scaled elliptic ﬂow for identified hadrons. (bootom) ratio
of difference between quark number scaled baryon $v_2$ and quark number scaled meson $v_2$ divided by the sum: $(B-M)/(B+M)$. 
Model predictions are also shown on the lower panel. Figure is taken from Ref. \cite{Sorensen:2007rk}.}
\label{v2fig}
\end{figure}

Thanks to coalescence models it was realized that elliptic flow provides also a way to understand
the hadronization mechanism in itself. In fact if hadronization goes through 
coalescence of constituent quarks the anisotropy at partonic level
propagates at hadronic level according to\cite{moln} :
\begin{equation}
v_{2,M}(p_T)\approx2v_{2,q}(p_T/2)\,,\,\,\,v_{2,B}(p_T)\approx3v_{2,q}(p_T/3).
\label{v2coal}
\end{equation}
This effect together with the shape of rise and saturation at parton level predicted by
the parton cascade \cite{moln02} for the partonic flow
allows to see the charateristic feature of coalescence that is a larger
elliptic flow for baryons, by a factor 3/2, and a similar
shift of the $p_T$ at 
which  the $v_2$ reaches the maximum value. These features are clearly seen
in the experimental data, as anticipated by Voloshin \cite{voloshin}, see Fig.\ref{v2fig}.
We notice that the coalescence mechanism tanslates the
hydrodynamical behavior into higher $p_T$ at hadronic level with the
effect being larger for baryons. This is seen also from the agreement of
hydrodynamics $v_2(p_T)$ for  mesons up to
$\sim 1.5$ GeV and for baryons up tp $\sim 2.5$ GeV .
As a general remark we like to point out that coalescence is the only known
hadronization mechanism that can lead to an enhancement of the partonic
elliptic flow. Someone may think that hydrodynamics gives the larger $v_2$, but
at fixed elliptic flow in the partonic stage coalescence can still enahance
the elliptic flow.
However, Eq.(\ref{v2coal}) is valid under the assumption of a uniform
phase space density \cite{Pratt:2004zq} and the approximation of a one dimensional
space in which only collinear and equal momentum partons recombine. 
Relaxation of these assumptions can in principle break  
``quark number scaling'' of the elliptic flow. For a discussion about effects
beyond the simplest (naive) coalescence, see the next Section.

In Fig. \ref{v2fig} (left) we show the scaled elliptic flow for pions and protons
obtained with our model once the quark $v_{2q}$ is fitted to reproduce $K$ elliptic flow,
then elliptic flow of $p,\bar{p},K,\Lambda$ is well reproduced \cite{greco2,Greco:2004yc}.
The calculations include the effect of resonance decays that is
another possible source of scaling breaking \cite{greco-res}, as dicussed
more in detail in the next Section.
The contribution to 
hadrons elliptic flow from minijet fragmentation is not included in this calculation. 
Its inclusion would lead to a universal hadron elliptic flow 
at momentum above $p_T \sim 6$ GeV \cite{fries}. 

In Fig. \ref{v2fig} (right), taken from Ref.\cite{Sorensen:2007rk}, the scaling of $v_2$ 
for various identified hadrons is shown
together with the relative baryon to meson ($(B-M)/(B+M)$) deviation from the naive scaling Eq.(\ref{v2coal}).
The circles are the experimental data, the dotted line is the prediction of our model \cite{greco-res},
and the solid line is the effect coming from higher Fock state in the hadron wave function 
discussed in Ref.\cite{Muller:2005pv}.
The agreement of our model with the data comes from a combination of the finite width of the hadron
wave function and the effects of resonance decays.

\begin{figure}[th]
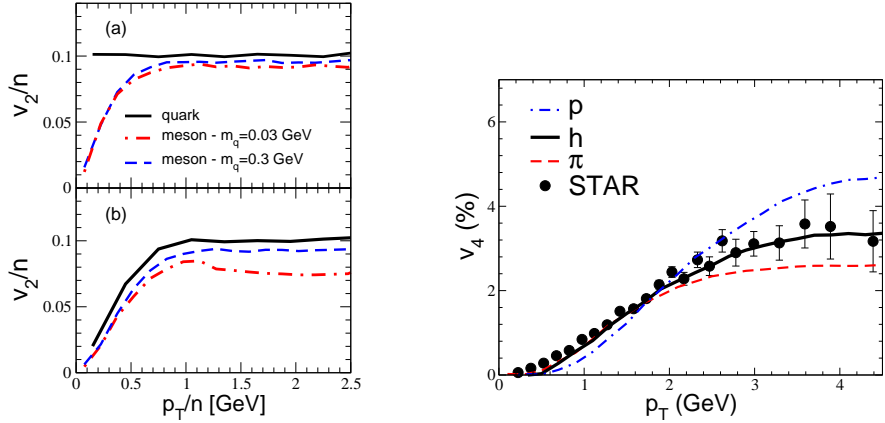

\begin{center}
\vspace{0.11in}
\includegraphics[height=2.2in]{v2-mass-dep2.eps}
\hspace{0.4 in}
\includegraphics[height=1.8in]{v4bm.eps}
\end{center}
\caption{Left: Scaled elliptic flows, the solid black line is the quark $v_{2q}$, the
dashed and dot-dashed line are the $v_2$ of mesons from coalesccence for two different
values of the quark mass $m_q$, see text for details. Right: $v_4$ in Au+Au collisions at 200 AGeV for charged (solid line), pion (dashed line)
and proton (dot-dashed line) evaluate from coalescence; circles are data taken from Ref.\cite{star-v4}.}

\label{v4bm}
\end{figure}

In Fig.\ref{v4bm} (left) it is shown the effect of quark mass on the scaling of $v_2$ at fixed
wave function width ($\Delta_p =$ 0.24 GeV). In panel (a) it is shown an ideal case 
in which the quark $v_{2q}(p_T)$ is flat, it is evident a strong violation of the scaling
at low $p_T$. The source of such a violation of the scaling is the full 6D phase space
that allows coalescence of non-collinear momentum at low $p_T$ where the boost effect is small.
In such a case the effect is nearly independent on the quark mass. 
In the lower panel (b) a more realistic case is shown, i.e. a $v_2(p_T)$ that rise and saturates
at $v_2=0.1$. In such a case the violation of the scaling at low $p_T$ is drastically reduced,
while at higher $p_T$ it is significantly mass dependent. We clearly see a violation of $\sim 25-30 \%$
if a mass $m_q= 0.03$ GeV is used. This is due to the fact that when the mass is small, even
if the wave function width is small in the rest frame of the hadron, particles with large
difference in the relative momenta can still coalesce due to boost effects. In such a case particles
at higher $p_T$ ($v_2 \sim 0.1$) can coalesce with particles at small $p_T$ ($v_2 \sim 0$) and this
effect is of course larger for smaller quark mass. 

It will be important to see experimentally where 
the scaling breaks down as a function of the beam energy but also as
a function of the rapidity. In principle one would expect that 
at very forward (backward) rapidity one reaches conditions that are comparable
to the midrapidity one at lower energy, therefore it could be that
a QGP phase is no longer dominant at large rapidity. 
Indeed we have seen that in the AMPT model the magnitude of $v_1$ and $v_2$ at 
$y \geq 3$ are consistent
with a pure hadronic matter dynamics \cite{v12-rap,chen-rap}.

The study of azimuthal anisotropy has been mainly limited to the 
elliptic flow (second harmonic of a Fourier expansion in the azimuthal angle) but a sizeable amount
of the fourth order momentum anisotropy has been predicted \cite{kolb-v4} and 
confirmed experimentally \cite{adams}. This provides the possibility
to further test the coalescence model, in fact also for the $v_4$ precise
relations between mesonic and baryon $v_4$ are expected \cite{greco-jam}, as for example:
\begin{equation}
\frac{v_{4B}}{v_{2B}^{2}}(3\, p_T)\approx
\frac{2}{3}\frac{v_{4M}}{v_{2M}^{2}}(2\, p_T)+\frac{1}{6}
\label{v4bm-form}
\end{equation}
Finite momentum spread and resonance decay do not change significantly
such a prediction, especially at $p_T > 1 \, GeV$ \cite{Greco:2004yc,greco-jam}.
For the details on the derivation of this and other  scaling relations
 also for odd harmonics we remand to Ref.\cite{coal-v4}.
In Fig.\ref{v4bm} (left) predictions for pion and proton $v_4$ are shown, the calculation
is done fitting $v_{4q}$ to the charged hadron $v_4$. Recent data \cite{Masui:2005aa} again confirm
the trend predicted by coalescence \cite{greco-jam,Greco:2004yc}.

\section{Beyond the simplest implementation}
The RHIC program has provided a remarkable evidence that coalescence 
of massive quarks supplies a simple and succesful model for hadronization
from a deconfined plasma. Nonetheless there are several aspects that are 
still problematic and act as a
stimolous to a deeper and advanced formulation of the coalescence process
in the context of hadronization.
Some of these problems are more serious if hadronization
via coalescence is applied to bulk production of hadrons.
Nevertheless, once coalescence is recognized to be the dominant mechanism
at intermediate $p_T$, it is appealing to investigate its behavior at low
momenta where the phase space is denser and coalescence probability increases.
Till now we have simply extended the approach at quite low momenta noticing 
that no striking contradiction with the experimental data are observed
which validates also the good agreement at intermediate $p_T$. 
Even if one should be aware of the accompanying problems, that we discuss
in the following, also the success of ALCOR in reproducing the total yields \cite{alcor}
has to be considered. Furthermore the reasonable
description of the experimental data on particle spectra also
at low $p_T$ in our approach is found also in a more recent based on
ALCOR/MICOR approach \cite{csiz04}. 
We want also to point out that an important
probe for the formation of QGP and the hadronization mechanism is given
by the charge fluctuations \cite{fluct}. First data 
\cite{mitch} 
agrees better with predictions from coalescence \cite{bialas}.
Furthermore charge fluctuation of the bulk are reproduced by a coalescence
hadronization mechanism if the number of quarks and antiquarks is of the order
of $dN/dy \cong 1300$ \cite{Nonaka:2005vr} (for most central collisions)
which is similar to the number used in the model
discussed here at intermediate $p_T$ and
to what is extracted by ALCOR model that is instead dedicated to the study of the multiplicities
\cite{levai-munster}. This shows again an internal consistency of coalescence models.

The simplest approach to hadronization by coalescence considers a uniform 
distribution of particles that combine if have
the same $p_T$ forming directly the stable hadrons. Of course this seems
too naive for a number of reasons that I try to list here together with some investigations
beyond this simple picture that have been or could be done in the next furture: 

\begin{enumerate}
\item \textit{Resonances} - The baryon/meson anomaly and moreover the
scaling of elliptic flow are directly applicable to stable hadrons, on the other
hand it is well-known that especially for pions there is a large feed-down from
resonance decays. 
In the model we have presented here and published originally in Ref. \cite{greco2}
resonances were already included.
\begin{figure}
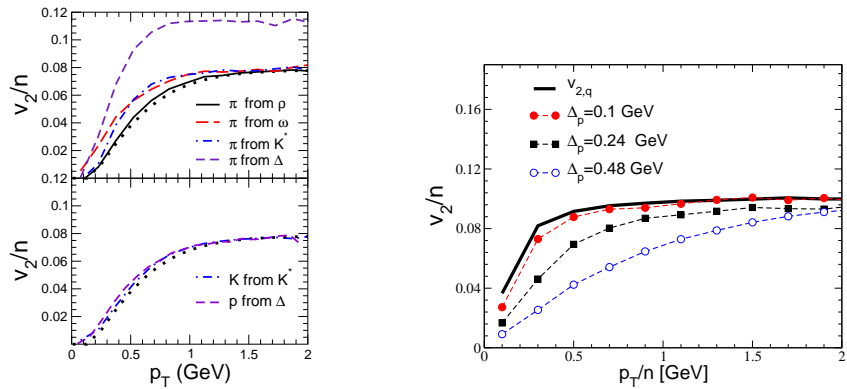

\begin{center}
\includegraphics[height=2.0in]{v2-reseff-2.eps}
\hspace{0.5in}
\includegraphics[height=1.7in]{v2-radial-therm-scal.eps}
\end{center}
\caption{Left: Scaled $v_2$ for (a) pions coming from different
resonane decays and (b) for K and p from $K^*$ and $\Delta$ respectively. Dotted line show
the underlying quark $v_{2q}$. Right: scaled $v_2$ for quarks (solid line) and for mesons for different
wave function width $\Delta_p$.}

\label{v2wf-res}
\end{figure}
As discussed more in detail in Ref.\cite{greco-res},
at intermediate $p_T$ their role is quite reduced and therefore their discard by other
groups is justified when looking at $p_T \simeq 3$ GeV. However our calculation shows
that when included both $p_T$ spectra and $v_2(p_T)$ have a better agreement with data 
at lower $p_T$.
In particular, we have studied how the elliptic flows of pions and other stable hadrons are
affected by decays of resonances, such as $ \rho \rightarrow 2\, \pi$, $\omega \rightarrow 3\, \pi$,
$ K^* \rightarrow K \,\pi$, and $\Delta \rightarrow p \, \pi$ \cite{greco-res},
see Fig.\ref{v2wf-res}. 
It turns out that particles like $p$, $\Lambda$, 
and $K$ from resonance decays have elliptic flows that are very 
similar to the directly produced ones. Therefore, the inclusion of 
resonances does not destroy the coalescence scaling of these 
stable hadrons. On the other hand, pions from the decay of $\omega$, 
$K^{\star}$, $\Delta$ show a significant enhancement of their elliptic 
flow at $p_T < 2$ GeV. The breaking of 
coalescence scaling due to resonance decays together with that
due to finite quark momentum spread (see next point) lead to a better 
agreement with available data as shown in Fig.\ref{v2fig}
where the decay of resonances is included 

However pions are not expected to be well described, but one has also
to consider that at intemediate $p_T$ the mass mismatch is less
relevant and at low momenta  most of the pions are coming from resonance
decays. This can be the reason behind the approximate
good description of pions in our model.
\cite{greco2,greco-res}. 

\item \textit{Wave function} - Our formalism described in Section \ref{formulation}
contains the full 3D phase space and a wave function with a finite width $\Delta_p$ of the order
of the Fermi momentum \cite{greco-res}.
The effect $\Delta_p$ is shown in Fig. \ref{v2wf-res} (right)
where we can see a significant increase of the violation of the scaling between
partons and mesons as a function of the width of the wave function. 
However with the same width used to reproduce the $p_T$ spectra 
the inclusion of a full 3D phase space and wave function
does not destroy but even improve the agreement with experimental data,
see Fig.\ref{v2fig} and the relative discussion \footnote{The $v_2(p_T)$ used here
is only schematic to show the effect of the violation that for a more realistic $v_2(p_T)$
is however reduced to about a $10 \%$ for $\Delta_p =$ 0.24 GeV.}.
An extreme ansatz for the wave function could be
considered, but if the wave function is flat in term of the 
relative momentum, the naive formulas Eq.(\ref{v2coal}) (right)
are strongly modified leading to a reduced difference between baryons
and mesons \cite{lin03} not observed experimentally. 
\begin{figure}
\begin{center}
\includegraphics[height=2.in]{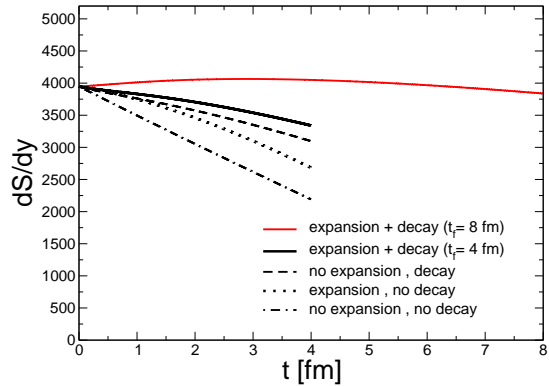}
\end{center}
\caption{Left: Evolution of the total entropy in $|y|<0.5$ associated to $u,d$ matter as a function of time,
for different options about the expansion and the decay of hadrons formed by coalescence. }
\label{entropy}
\end{figure}

\item \textit{Energy and Entropy Conservation} - Another issue related to 
Eq.(\ref{coal1}) is the energy conservation.
In the present approach on shell massive quarks are recombined into one
particle respecting momentum conservation, but not energy conservation. 
Of course this is a drawback of the 
simplest implementation while recombination processes always happen in nature
involving a third particle or off-shell recombining particles. 

In our model pions
come mainly from $\rho$ decay hence the energy violation is less serious even 
at low $p_T$, at the same time both the $p_T$ distribution and the
elliptic flow get closer to data thanks to a shift to lower $p_T$.
For example for the $v_2$ of pions gets closer to a scaling with
$m_T - m_0$ as found experimentally \cite{Adare:2006ti}, which can be seen
as an indirect effect of energy conservation.
More recently energy conservation in the coalescence process 
has been investigated by Ravagli and Rapp \cite{Ravagli:2007xx}
using a formulation of coalescence from the Boltzmann collision integral.
They find that if energy conservation is implemented then the $v_2$ scaling 
as a function of $p_T$ moves toward a scaling versus $m_T - m_0$.
Another important development is the inclusion of a mass distribution for the quarks
as an effective way for including in-medium interaction \cite{Zimanyi:2005nn}, this is another way
to allow for both momentum and energy conservation and leads to a fairly good
agreement with the data, for more details see the contribution of T. Bir\'{o} to these
Proceedings.

An issue related also to energy conservation is the entropy conservation. Coalescence 
reduces the number of particles by about a factor two (if resonance production is not
taken into account), this rises suspects of a strong decrease
of entropy. Of course entropy is not the number of particle, but 
depends also on the degeneracy of the two phases and on the mass
of the particles. For example converting one gluon into one pion with
the same mass leads to a huge entropy violation. In other words a large
chemical potential is enforced by number conservation causing an entropy
decrease unless at the same time there is a large volume expansion. 

In Fig. \ref{entropy} it is shown the evolution of $dS/dy$ (for $|y|<0.5$ as a function of time if one assumes
that coalescence takes place during a mixed phase of 4 fm (and 8 fm upper line). 
during which the volume expands with $\beta=0.5$ and quarks are gradually
converted into hadrons which are allowed to decay according to their width.
For the full calculation (expansion plus decays) a decrease of about 15 $\%$ is found, but
in our coalescence model there is also a decrease in the
energy by about $~ 15\%$ ($S=(E+ln Z)/T $, where $Z$ is the partition function).
Therefore if coalescence is implemented taking care of the
energy conservation probably entropy can also be conserved.
For the multiplicity of various hadrons the calculation is based on the results from our coalescence model
(but only $u,d$ matter is considered).
The important role of the decays and the moderate role of the expansion are shown
by the different cases considered, see Fig. \ref{entropy}.
We also see that assuming a longer mixed phase ($t_f=8$ fm), and therefore a larger volume at the
end entropy can be conserved even in such a simple approach.
In addition one should also take into account the interaction among quarks; it has been shown
using the lQCD equation of state for an isentropically expanding fireball that the
evolution of the effective number of particles reduces significantly
around the crossover temperature \cite{Biro:2006sv}. 
This of course helps to solve the entropy problem inherent to
quark coalescence, as pointed out also by Nonaka et al. \cite{Nonaka:2005vr}.

\item{\textit{Space-momentum correlation and $v_2$ scaling}} - 
The constituent quark number scaling of the elliptic flow was derived by Voloshin
and Molnar assuming that the x-space can be decoupled from p-space and integrated out.
This means that the scaling has been explicitly demonstrated only if the coalescence
probability is omogeneous in space and if space and momentum
are not correlated. A detailed discussion on effects coming from space and momentum correlation
as a source of scaling violation can be found in Ref.\cite{Pratt:2004zq} together
with classes of distribution that lead to an approximate scaling.
Another study on the effect of phase space distribution can be found in Ref.\cite{Greco:2005jk}.

Results presented here with our coalescence model do not assume the coalescence scaling
and show that if radial flow space-momentum correlation (but still in a uniform
fireball) is included the scaling is still recovered \cite{greco2}. 
An important tool of investigation in this context is supplied by parton cascade 
studies that calculate the time evolution of the phase space. 
A first investigation by Molnar finds that the scaling between baryons and mesons
still persists even if a strong violation is found respect to the $v_2$ at quark level
\cite{moln04}.
However it is not clear the dependence on the freeze-out criteria, on the wave function
width and on the interplay with
indipendent fragmentation mechanism as sources of scaling breaking, see also
D. Molnar contribution to these Proceedings. 

Finally it is clear that a better understanding of HBT measurements can supply fundamental
information on this issue that is the potential main source of scaling violation
and therefore should be investigated in deeper detail.
At LHC the different dynamical evolution may lead to an even stronger scaling violation.

\item \textit{Jet-like correlation} -
The last but not the least issue is related to the correlation among
hadrons in the hadronization process. At variance with the other
issues this is the one more relavant at intermediate $p_T$.
At RHIC it has been possible
to measure the correlation between hadrons at momentum $p_T^{trig} >$4-6 GeV 
and the 
associated particles with momenta $p_T < p_T^{trig}$ as a function of the
azimuthal angle respect to the trigger particle \cite{star-corr,phenix-corr}. 
Such a kind of measurement has shown that hadrons at intermediate $p_T$ come with 
associated particles at lower momentum in the opposite direction. 
In the seminal papers
\cite{greco,greco2} we have taken into account the possibility of coalescence
between thermal partons and minijets and this was a primitive way to propose
the idea that at intermediate $p_T$ particle from coalescence could follow
the di-jet correlation.  
A formal framework for the effect of quark correlations has been  
studied by R. Fries et al. \cite{Fries:2004hd} and R. Hwa and C.B. Yang \cite{hwa-corr}.
They have shown that coalescence goes along with particle
angular correlation, even if more stringent tests
are currently under investigation. 
Anyway the main problem remains the study of the origin of correlation at parton
level which asks for a transport approach, a first explorative study has
been performed by Molnar \cite{Molnar:2005wf}.
\end{enumerate}

In the future it is desireable to have a deeper theoretical investigation
of coalescence, in fact it is easy to predict that such a kind of
processes will be even more dominant at the Heavy-Ion experiments at the
Large-Hadron-Collider (LHC) at CERN, see also P.Levai contribution
to these Proceedings. In fact hard processes will be largely
dominant together with a larger jet quenching and a thermal source with
larger radial flow which should lead to an overwhelming production of hadrons from
coalescence in a wider momentum range \cite{fries-LHC,levai-munster}.
Therefore the features of more sophisticated models will have wide chances 
to undergo many experimental tests at both the LHC and RHIC programs.

\section{Phase Space Coalescence for heavy quarks}
The coalescence model has been also applied to study open and hidden charm and bottom
production. For the evaluation of charmonia yield the idea of coalescence \cite{grandchamp}
or recombination at phase boundary \cite{charm-stat} has been investigated by several authors 
as a competing effect to charmonia suppression.
We have suggested instead that the dominance of coalescence mechanism in a wide
range of momenta can be applied also for the study of the D and B mesons \cite{greco-c,vanHees:2005wb,van Hees:2007me,van Hees:2007mf}
that are the principal tool for investigating heavy quark interaction 
and thermalization in the
QGP. In a first paper we have shown that such an approach allows to estimate both
spectra and elliptic flow of the D mesons and that this offers the
possibility to relate $ J/\Psi$ and D meson spectra if both come
mainly from coalescence \cite{greco-c}.

Surprisingly, data from the
Relativistic Heavy-Ion Collider (RHIC) for single electrons ($e^{\pm}$)
associated with semileptonic $B$ and $D$ decays in semi-central Au-Au
collisions exhibited a $v_2$ of up to 10$\%$~\cite{Kelly:2004,adare07},
indicating substantial collective behavior of charm ($c$) quarks
consistent with the assumption of a $c$-quark $v_2$ similar to the one
for light quarks, apart from a $p_T$ shift due to radial-flow effects
\cite{greco-c}. In addition, the nuclear suppression factor was
found to be comparable to the pion one, $R_{AA}$$\simeq$0.3
\cite{Adler:2005xv,abelev07}.

On the other hand for heavy quarks (HQ) the predicted amount of energy-loss has been shown to be insufficient 
(at variance with the light quark case) to
account for the observed non-photonic single electrons nuclear suppression (small $R_{AA}$) and
for a strong degree of collectivity (large $v_2$) \cite{armesto05}. 
Therefore for HQ the challenge is mainly
 the understanding of the quark in-medium interaction, even if the acquired knoweledge
on hadronization mechanism from light quarks plays a significant role.

Based on lattice QCD (lQCD)
which suggests a resonance structure in the meson-correlation function at
moderate temperatures \cite{latt-charm}, an effective model for heavy-light quark
scattering via $D$ and $B$ resonances has been suggested.
In Ref.\cite{vanHees:2004gq} it was found that resonances cause a reduction of about
a factor of three in the thermalization time respect to pQCD estimates.
To relate such a microscopic description to the single elctron data, 
we employ a Fokker-Planck approach for $c$ and $b$ quarks in the QGP based on elastic
scattering with light quarks via $D$- and $B$-meson resonances with a
width $\Gamma$$=$$400$-$750$~MeV (supplemented by perturbative
interactions in color singlet channel)~\cite{vanHees:2004gq}. Heavy-quark (HQ) kinetics in the
QGP is then treated as a relativistic Langevin process~\cite{vanHees:2005wb}:
\begin{equation}
\frac{\partial f_Q}{\partial t} = \gamma \frac{\partial (p f_Q)}{\partial p}
+ D_p \frac{\partial^2 f_Q}{\partial p^2} \ , 
 \end{equation}
where $f_Q$ is the HQ phase space distribution, $\gamma$ and $D_p$
are the corresponding drag and (momentum) diffusion constants which
determine the approach to equilibrium and satisfy the Einstein relation, $T=D_p/\gamma M_Q$.
The medium is modeled by a spatially homogeneous elliptic thermal
fireball which expands isentropically.
Finally the hadronization is treated by a coalescence (see Eq.(\ref{coal1})) plus fragmentation approach
with the distribution of HQ that undergoes a fragmentation process evaluated as
$f_{c,b}(p_T)*[1-P_{c,b\rightarrow(D,\Lambda_c),(B,\Lambda_b)}(p_T)]$, where $P_{c,b\rightarrow (D,\Lambda_c),(B,\Lambda_b)}$ is the probability
for a HQ to coalesce according to Eq.(\ref{coal1}).
\begin{figure}[th]
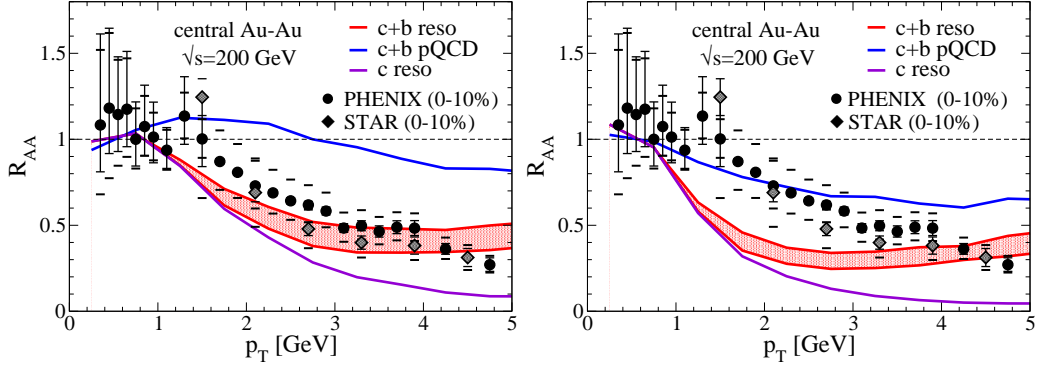

\includegraphics[height=1.9in]{raa_e_cent-therm.eps}
\includegraphics[height=1.9in]{raa_e_cent_fragonly-qm05data.eps}

\caption{Nuclear suppression factor for single-electrons, including coalescence and fragmentation
at hadronization (left panel) and only with fragmentation (right panel), see text.}
\label{raa}
\end{figure}
Results for Au+Au at 200 AGeV from the Langevin simulation
including hadronization by coalescence+fragmentation (left) and fragmentation only (right) 
are shown in Fig.\ref{raa} together with experimental data \cite{Adler:2005xv,bielcik06}.
It is clear that elastic scattering in a pQCD scheme is insufficient to account for the 
small $R_{AA}$ and large $v_2$ (see Fig.\ref{fig4}) independently on the hadronization scheme applied. 
The red band shows 
the full calculation with $c,b$ quarks that scatter in the presence of hadronic-like 
resonances with a width $\Gamma \sim 0.4-0.75$ GeV. We notice
that contamination of single-electrons from B decay is significant already at $p_T \sim 2$ GeV
(corresponding to a cross-point between $c$ and $b$ spectra around $4-5$ GeV in agreement
with FNNLO calculation \cite{Cacciari:2005rk}). Therefore
it is necessary to include the $B$ mesons (despite the inherent uncertainties in the
b/c ratio) to draw any conclusion on the interaction processes
behind the experimental results.

\begin{figure}[th]
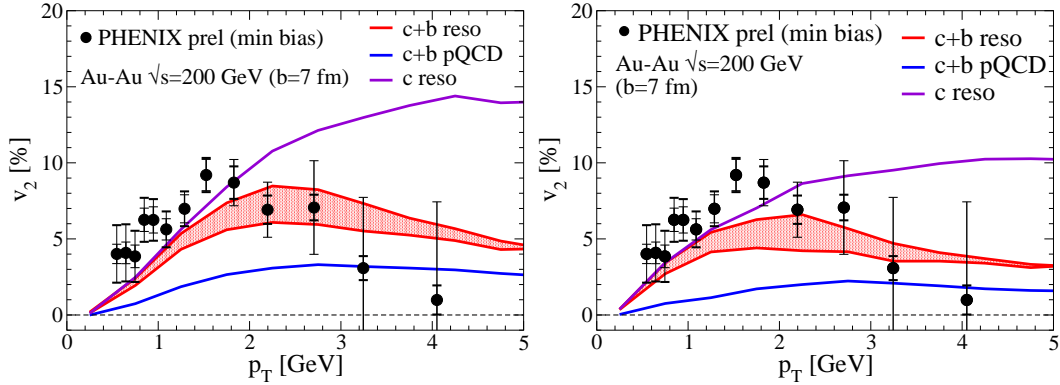

\includegraphics[height=2.in]{v2_e_MB_therm-av-qm05data.eps}
\includegraphics[height=2.in]{v2_e_MB_delta_frag_only-qm05data.eps}

\caption{Elliptic flow for single-electrons, including coalescence and fragmentation
at hadronization (left panel) and only with fragmentation (right panel), see text.}
\label{fig4}
\end{figure}
Comparing the band in the left and right panel of Figs. \ref{raa} and \ref{fig4}, we can
appreciate the effect of coalescence in the increasing of $R_{AA}$ at $p_T \sim 1-4$ GeV and the simultaneous 
increase of the elliptic flow $v_2$, see Fig.\ref{fig4} \cite{greco2,greco-c}. Therefore coalescence mechanism
reverts the usual correlations between $R_{AA}$ and $v_2$, and allows for a reasoble agreement
with the experimental data of both $R_{AA}$ and $v_2$. 

Very recently within a Brueckner many-body scheme in medium T-matrices for heavy-quarks scattering
off light quarks \cite{van Hees:2007me} have been evaluated starting from lQCD potential.
Therefore the existence and width of $D$ and $B$-like resonances are no longer assumed like in the 
effective lagrangian but extracted from lQCD.
It is found even a better agreement with the data \cite{van Hees:2007me}. Even if inherent uncertainties
in the extrapolation of the potential have to be evaluated in the next future, this approach constitutes
a promising tool to connect the observables to the information from lQCD.
Furthermore we note that if resonant scattering with increasing strength at decreasing temperature
is the dominat interaction channel this would lead to a natural merging into a quark-coalescence.
Therefore heavy quarks may allow a coherent description between in-medium interaction that
drives the thermalization process and 
the subsequent hadronization.

We also mention that LHC will play an important role to validate such a picture.
In Ref. \cite{van Hees:2007mf} predictions for LHC have presented. They are
quantitatively rather similar to our RHIC results~\cite{vanHees:2005wb},
due to a combination of harder initial HQ-$p_T$ spectra and a decrease in interaction strength in the early
phases where non-perturbative resonance scattering is inoperative.
Therefore we conclude that if at RHIC the dominant contribution to HQ
interactions are hadron-like resonances, at LHC we should observe an
$R_{AA}$ and $v_2$ pattern similar to RHIC.

Finally we point out that in the next future we can look not only at the quarkonia yields,
but scrutinize their $p_T$ which allows to check the self-consistency
between D and B mesons and $J/\Psi$ and $\Upsilon$ quarkonia that in
the QGP should be related by the unique underlying HQ distribution. 
Therefore phase space coalescence will provide a much deeper insight into the long-standing issue
of $J/\Psi$ suppression and regeneration.

\section{Summary and Conclusions}
The first stage of RHIC program has shown clear signs of modification of the hadronization
mechanism  respect to $pp$ collisions in the light quark sector.
There are several evidences that hadronization proceeds through coalescence of massive quarks
close in phase space. The baryon-to-meson ratio enhancement observed at intermediate $p_T$ and the
scaling of the elliptic flow with the number of valence quarks, are robust effects against
improvements of the naive coalescence picture:
inclusion of resonances, finite width of the wave function, effective quark mass
distribution, gluons in the higher Fock staes of the wave function, energy
conservation ...
All these studies that improve the original coalescence picture from a theoretical
point of view lead generally
to an even better agreeement with the data. The main open question
remains the development of a dynamical coalescence model that
can shed light on the issue of space-momentum correlation and jet-like correlations
that could be significant sources of deviation from naive coalescence.

The knoweledge of hadronization for light quarks seems to play a role
also in the new challenges posed by heavy-quark probes.
Here the main issue is the dominant interaction mechanism and its relation to the microscopic
structure of the QGP. The role of coalescence is mainly a modification of the
correlation between $R_{AA}$ and $v_2$.
Finally, coalescence in the heavy quark sector can provide an inherent consistency between 
the in-medium interaction of HQ and the subsequent hadronization 
considering that a pole in the HQ propagator above
$T_c$  can be viewed as a precursor of coalescence.

\section{Acknowledgments}
I was very pleased to participate to the Workshop in honor of Prof. J. Zimanyi whom scientific
activity has fortunately influenced mine through the collaboration with P. Levai that I thank
for his kind invitation.

I would like also to thank several collaborators during different phases of the work presented:
C.M. Ko, L.W. Chen, P. Levai, H. Van Hees, R. Rapp and I. Vitev.
I am grateful to G. Ferini for her careful reading of the manuscript.

\end{document}